\documentclass[figures,nocite]{epl}

\shorttitle{Experimental evidence of the dependence of spin
tunneling...}

\shortauthor{F. Torres {\it et al.}}

\title{Experimental evidence of the dependence of spin tunnelling on the concentration of dislocations in Mn$_{12}$ crystals}

\author{F. Torres \inst{1}, J. M. Hernandez  \inst{1}\thanks{E-mail: \email{jmhdez@ubxlab.com}}, E. Molins     \inst{2}, A. Garcia-Santiago \inst{1} and J. Tejada \inst{1}}
 \institute{
     \inst{1}Dept. F\'\i sica Fonamental, Univ. Barcelona. Diagonal
647. 08028 Barcelona. Spain. \\
     \inst{2}Institut de Ci\`{e}ncia de Materials de Barcelona - CSIC.
Campus UAB. 08193 Cerdanyola. Spain}


\begin{document}
\maketitle

\begin{abstract}
We present experimental results on resonant spin tunnelling in a
single crystal of Mn$_{12}$-2Cl benzoate with different
concentration of dislocations. The time evolution of the
magnetisation follows the stretched exponential, $M(t)\propto
\exp[-(t/\tau)^{\beta}]$ over a few time decades. The values of
$\tau$ and $\beta$ deduced from experiment have been used to
determine the concentration of dislocations before and after the
cooling-annealing process, using the algorithm recently suggested
by Garanin and Chudnovsky.
\end{abstract}

The crystals of molecular clusters are soft and fragile. They must
contain a large number of dislocations which rearrange atoms in
their immediate vicinity. Dislocations are formed by the frozen in
lattice disorder when crystallization takes place. In the case of
molecular clusters, the dislocations can also be introduced by
performing thermal cycles which produce dilations and compressions
of the lattice. Many techniques are now available for the study of
dislocations. The two most important techniques are X-ray
diffraction and transmission electron microscopy.

After five years of experiments \cite{1,2,3,4,5,6} and theoretical
studies \cite{7,8,9,10,11} a new theoretical approach \cite{12}
has been suggested which may fully explain the mechanism of
thermally assisted resonant tunnelling in crystals of Mn$_{12}$
molecular clusters. Chudnovsky and Garanin have theoretically
shown \cite{12,13} that spin tunnelling in Mn$_{12}$ clusters must
be dominated by dislocations. Some of the theoretical suggestions
of Chudnovsky and Garanin have been recently confirmed in
experiments \cite{14,15,edb}. In this paper we report results of
our study of the effect of crystal defects on the tunnelling
process. For this purpose, we have compared the results of quantum
magnetic relaxation experiments on a single crystal of Mn$_{12}$-
2Cl benzoate (Mn$_{12}$Cl) before and after suffering thermal
shocks.

At low temperature the intramolecular exchange interactions render
an effective spin $S$=10  for this molecular cluster. The
Mn$_{12}$Cl crystallizes in an orthorhombic structure, a = 2.275
nm, b = 1.803 nm and c = 1.732 nm with two molecules per unit cell
\cite{16,17}. The magnetic core and the local symmetry of each
molecule are identical to the case of Mn$_{12}$Ac. However, in
contrast to Mn$_{12}$Ac, the magnetic easy axes of Mn$_{12}$Cl
molecules lie alternatively on the direction ($011$) or
($0\overline{1}1$), being nearly perpendicular to their nearest
neighbors. A fresh single crystal of Mn$_{12}$Cl clusters was
first characterized by X-ray diffraction. Then we performed low
temperature magnetic relaxation experiments. After this we cycled
the temperature of the crystal between 80 K and 300 K by
introducing it alternatively in liquid nitrogen during five
minutes and in water during five minutes. This cooling-annealing
process was repeated four times after which the crystal was again
characterized by both X-ray diffraction and magnetic relaxation. A
second cooling-annealing treatment, followed by X-ray and magnetic
characterization measurements, was also performed. Our idea is to
combine the thermal treatments with the defects induced during the
X-ray characterization to increase the number of dislocations.

The crystal was mounted on the top of a glass capillar, glued at
the prism base with its long dimension, ($012$) direction,
oriented along the capillar. The crystal was kept on the capillar
along all the performed thermal shocks, diffraction and
magnetisation experiments.

The thermal shocks generate a large temperature gradient in the
crystal producing radial and tangential tensions that favor the
propagation of dislocations across the crystal, probably starting
at point defects frozen during the growing of the crystal. The
extension of these dislocations by the whole crystal converts the
initial single crystal in a multidomain crystal being each element
of it slightly misaligned with respect to its neighbors. This is
what is known as a mosaic crystal and it is well known that as the
crystals contain more and more dislocations greater is the
mosaicity. The amount of misalignment is related with the widening
of the Bragg peaks along the $\omega$ \cite{23,24,25} axis and in
the case of our single crystal after the thermal shocks, is of
some tenths of a degree. Due to this low value, the crystal is
still considered as a single crystal but with a larger mosaicity.
The shape and size of a Bragg peak, except for instrumental
parameters, depend on the shape and size of the crystal, on the
spectral composition of the X-ray beam and on the mosaicity of the
crystal. It is also known that the lack of monochromaticity
elongates the peak along the $\omega-2\theta$ direction
\cite{18,19,20}. This effect together with mosaicity widening
leads to an idealized hexagonally shaped diffraction peak. In our
experiments low angle peaks were selected in order to reduce the
monochromaticity effect.

\begin{figure}
\onefigure[width=4.5in]{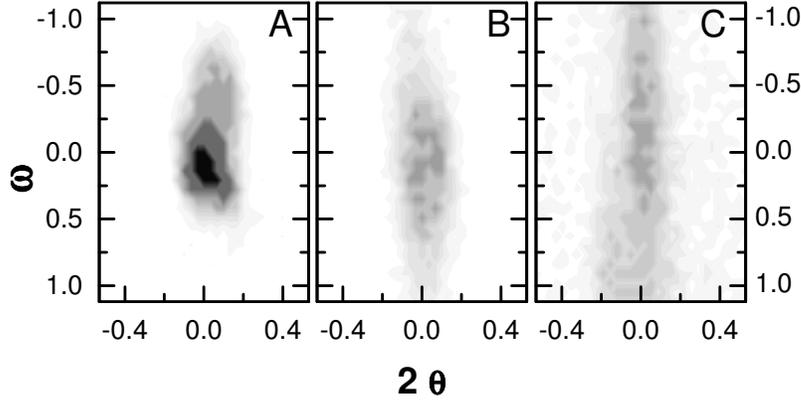} \caption{$\omega -\theta$ plot
of the ($\overline{2}22$) reflection of the Mn$_{12}$Cl crystal:
before (A), after one heat treatment (B) and after two (C). They
clearly show the enlargement of the peak along omega direction due
to the increase of the mosaicity after the heat treatments.}
\label{f.1}
\end{figure}

After a random searching of diffraction peaks at $\theta$ angles
larger than $10^{\mathrm{o}}$, the spots were indexed to check
their single crystal origin, the crystal orientation and to get
their cell parameters by using a four-circle single-crystal X-ray
diffractometer (Enraf-Nonius CAD4, MoKa radiation). The
reflections ($\pm2,\pm2,\pm2$) were selected as they were low
angle and intense. After centering, bi-dimensional profiles of
these peaks were recorded at constant steps on $\omega$ and
$\theta$. The same profiles were also measured after the thermal
treatment. See the two peaks on Figure \ref{f.1}.

Although a precise analysis of diffraction spots shape is
complicated, it is easier in the current case as we are comparing
the same reflections on the same crystals, we can attribute all
the changes to variations of mosaicity. All the measured
reflections showed a widening of the peaks along the $\omega$
direction after the thermal treatment, while keeping similar
widths in the $\theta$ direction. One of those reflections has
been arbitrarily selected and depicted in Figure \ref{f.1}. The
$\omega$ widening even overcomes the predefined scan width, as can
be clearly seen in the Figure \ref{f.1}. A flattening of the peaks
occurs, reducing their maximum intensity by a factor of about 3
due to the $\omega$ spreading out. In Table \ref{t0} appear the
values of the $\omega$ widths, $\Delta \omega$, obtained by
fitting the peaks of Figure \ref{f.1} to a 2D lorentzian function.
It is clear that $\Delta \omega$ increases with the heat
treatments. Assuming that the concentration of dislocations is
proportional the $\omega$-widening, it may be concluded that the
number of dislocations existing in the Mn$_{12}$ single crystal
has increased by near an order of magnitude with the thermal
processes.

\begin{table}
\centering \caption{$\omega$-widths of the ($\overline{2}22$)
reflection peaks obtained by fitting them to the 2D lorentzian
$\frac{A}{1+\left ( \omega / \Delta \omega \right )^2 +\left (
\theta / \Delta \theta \right )^2}$.} \label{t0}
\begin{tabular}{lc}
 H.T.& $\Delta \omega$  ($\times 10^{-3}$ degree)\\ \hline
 Before & $222 \pm 6 $ \\
 After 1st & $479 \pm 12$  \\
 After 2nd & $1470 \pm  60$ \\
\end{tabular}
\end{table}

The energy barrier that separates the spin states  $m > 0$ and $m
< 0$ in Mn$_{12}$ molecular clusters derives from the Hamiltonian
\begin{equation}
{\cal H} = - AS^{2}_{z} - BS^{4}_{z} + ½ C(S_{+}^{4} + S_{-}^{4})
+ {\cal H}_{dip} + {\cal H}_{hf},
\end{equation}
where $A = 0.38 (1) cm^{-1}$ and $B = 8.2(2) \times 10^{-4}
cm^{-1}$ \cite{18,19,20} and $C = 3 \times 10^{-5} cm^{-1}$
\cite{18,19,20,21,22}. The first two terms of the spin Hamiltonian
generate spin levels $S_{z}$ inside each well and the $C$ term
corresponds to the fourth order anisotropy term which contains
spin operators that do not commute with $S_{z}$ causing,
therefore, tunnelling. ${\cal H}_{dip}$ and ${\cal H}_{hf}$ are
due to magnetic dipole fields and hyperfine interactions,
respectively, which also cause tunnelling. When an external field
is applied along the easy axis of the molecules producing the
crossing of energy levels, the shortcut between degenerate $m$ and
$-m$ levels, at the resonant fields, has been interpreted as due
to the tunneling. In the presence of tunnelling, the activation
energy is, therefore, at the $m$ tunnelling level instead of that
at the top of the energy barrier. But with only these terms in the
Hamiltonian it is not possible to get a fully quantitatively
explanation for all the experimental facts. Very recently,
Chudnovsky and Garanin \cite{12,13} have suggested the existence
of a new term in the Hamiltonian, ${\cal H}_{me} =
E(S_{x}^{2}-S_{y}^{2})$, which is due to the magnetoelastic
coupling and mostly causes the spin tunnelling in the Mn$_{12}$
clusters. That is, this second order anisotropy term which is
prohibited in a perfect crystal of Mn$_{12}$, is always present in
real crystals due to the lattice distortions associated with the
presence of dislocations.

\begin{figure}
\onefigure[width=4in]{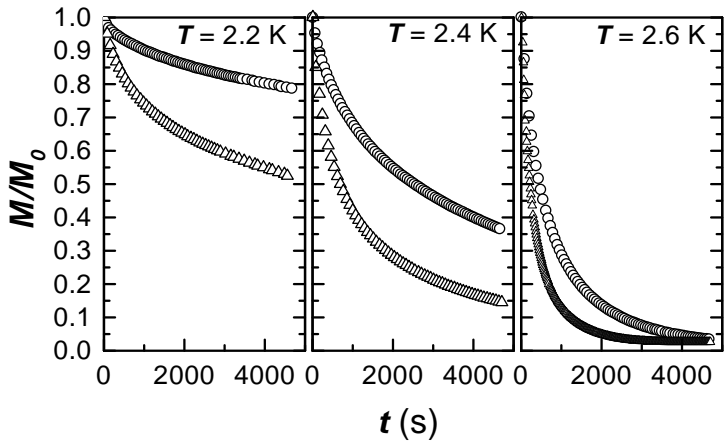} \label{f.2}
 \caption{Magnetic
relaxation of a Mn$_{12}$Cl single crystal from saturation
measured at different temperatures, before (circles) and after two
heat treatments (triangles).}
\end{figure}

The magnetic relaxation experiments were performed as follows: the
single crystal is cooled in the presence of a 3 T field from above
the blocking temperature and at the desired lower temperature the
magnetic field is switched off. The variation of the magnetisation
with time, $M(t)$, is then recorded. Figure \ref{f.2} shows $M(t)$
before and after the second cooling - annealing process at three
different temperatures. These curves have been fitted, within the
entire time interval, by a stretched exponential function
$M(t)\propto \exp[-(t/\tau)^{\beta}]$. In Table \ref{t1}, we list
the values for $\tau$ and $\beta$ deduced from the fitting
procedure. We have found that the values of both $\tau$ and
$\beta$ decrease when increasing the concentration of
dislocations. Both $\tau$ and $\beta$ have been found to decrease
when lowering the temperature of the relaxation experiment.

\begin{table}
\caption{Parameters obtained from the fitting of the relaxation
data to a stretched exponential function for different
temperatures and heat treatments.} \label{t1}
\begin{center}
\begin{tabular}{llcc}
$T$ (K) & H.T. & $\tau (s)$ & $\beta$ \\ \hline
2.2& Before & $69 \times 10^{3}$ & 0.54 \\
 ~& After 1st & $57 \times 10^{3}$ & 0.51\\
 ~& After 2nd & $10 \times 10^{3}$ &0.43\\
 \hline
2.4& Before & $46 \times 10^{2}$ & 0.59 \\
 ~& After 1st & $39 \times 10^{2}$ & 0.53\\
 ~& After 2nd & $10 \times 10^{2}$ &0.46\\
 \hline
2.6& Before & $690$ & 0.66 \\
 ~& After 1st & $700$ & 0.65\\
 ~& After 2nd & $240$ &0.54\\
\end{tabular}
\end{center}
\end{table}

The question now is to use, if its possible, the variation of the
values of $\tau$ and $\beta$ to get a quantitative estimation of
the variation of the density of dislocations with the thermal
process. That is, the curves of Figure 2 correspond to the
magnetisation relaxation associated with the tunnelling events
contributing to the resonance at zero field and are very sensitive
to the concentration of dislocations. The reason for that lies in
the fact that the value of the transverse anisotropy causing
tunnelling for each molecule is very much dependent on the
location of the molecule with respect to the dislocation. As
Chudnovsky and Garanin have demonstrated \cite{12,13} the
tunnelling probability of different spin levels overlap for
different molecules. That is, the tunnelling magnetisation at each
time is associated with the broad spectrum of molecules located at
different distances from the core of a dislocation.

A few observations are in order before discussing the model we
propose to explain our data: a) The molecules close to the nucleus
of dislocations are those relaxing faster and their number
increases with the density of dislocations, b) As the temperature
of the relaxation experiment increases, more and more molecules
located far away from the dislocations contribute to the
relaxation, c) Only in the absence of dislocations the relaxation
should be pure exponential, d) due to dislocations there is a
distribution of transversal anisotropy $E$ values \cite{12,13}
associated with different locations of molecules with respect to
the core of a dislocation. This distribution, $f(E,c)$, may be
well represented by a Gaussian which center and width depend on
the concentration of dislocations, $c$. The fact that the
tunnelling rate depends strongly on $E$ values suggest that there
is a distribution of transition probabilities.

That is, the amount of relaxing magnetisation at each time is
written as

\begin{equation}
M(t) \propto \int _0 ^\infty  \mathrm{e}^{-\Gamma(E,T) t}f(E,c)
\mathrm{d}E \label{m(t)}
\end{equation}

\noindent where $\Gamma(E,T)$ is the effective rate of relaxation
for those clusters with perpendicular anisotropy $E$. In order to
compute this parameter we have used the expression given by
Friedman \cite{Friedman} \footnote[1]{This expression allows us to
consider the transition probabilities through each level, fact
that must be introduced to explain the thermally assisted nature
of the resonant tunnelling phenomena observed in the kelvin
regime.}

\begin{equation}
\Gamma(E)= \Gamma_0 \exp (-E_0/T)+ \sum_{m=1}^{10} \frac{\Gamma_0
}{1 +   \frac{\Gamma_0^2}{ 2 \Delta_m ^2} } \exp (-E_m/T)
\end{equation}

\noindent where

\begin{equation}
\Delta_m= \frac{2 D}{\sqrt{(2 m -2)!!}} \frac{(S+m)!}{(S-m)!}
\left(\frac{E}{2 D}\right)^{m}
\end{equation}

\noindent is the tunnelling splitting for the level $m$,
$\Gamma_0$ is the attempt frequency, $T$ is the temperature, and
$E_m$ is the energy of level $m$. Although the analysis of
Chudnovsky and Garanin was performed for the uniaxial crystal of
Mn$_{12}$-Acetate, the above formulas are rather general and can
be equally applied to Mn$_{12}$Cl crystals.

Solving equation (2) we have seen that for a wide range of $T$ and
$c$ values, the resulting magnetic relaxation follows very well
the stretched exponential function. By direct comparison of the
values of the parameters $\tau$ and $\beta$ deduced theoretically
with those obtained from the experiment, we have estimated the
concentration of dislocations of the crystal before and after the
heat treatments. The results are summarized in Table \ref{t2} and
show clearly that the concentration of dislocations increases with
the heat treatments.

\begin{table}
\caption{Estimated concentration of dislocations obtained from
comparison of experimental results with the results of the
calculations. Ratio between the $\omega$-width and the
concentration of dislocations.}\label{t2}
\begin{center}
\begin{tabular}{lcc}
 H.T.& $c$ ($\times 10^{-4}$)&$\Delta \omega / c$ \\ \hline
 Before & $3.0 \pm 0.5 $ & $740 \pm 140$\\
 After 1st & $6.0 \pm 1.0 $  & $700 \pm 150$\\
 After 2nd & $20 \pm  4$ & $740 \pm 180$\\
\end{tabular}
\end{center}
\end{table}

Assuming that the $\omega$-width, shown in Table \ref{t0}, is
proportional to the concentration of dislocations, the ratio
$\Delta \omega / c$ should be a constant. In Table \ref{t2}, we
present the values of this ratio for the different thermal
treatments. It can be seen that the ratio $\Delta \omega / c$
remain constant with the heat treatments.

In conclusion, combining the results of X-ray diffraction and
magnetic relaxation we have show quantitatively that the tunneling
probability increases when introducing dislocations in the matrix.

\acknowledgments This work has been supported by the EC
Commission, project number 4538 and by the Generalitat de
Catalunya (Grant 1999SGR-0205).

\end{document}